\documentclass{article}

\usepackage{arxiv}

\usepackage[utf8]{inputenc} 
\usepackage[T1]{fontenc}    
\usepackage{hyperref}       
\usepackage{url}            
\usepackage{booktabs}       
\usepackage{amsfonts}       
\usepackage{nicefrac}       
\usepackage{microtype}      
\usepackage{lipsum}		
\usepackage{graphicx}
\usepackage{natbib}
\usepackage{doi}

\title{Assessing Cerebellar Disorders With Wearable Inertial Sensor Data Using Time-Frequency and Autoregressive Hidden Markov Model Approaches}

\author{ \href{https://orcid.org/0000-0003-1312-6473}{\includegraphics[scale=0.06]{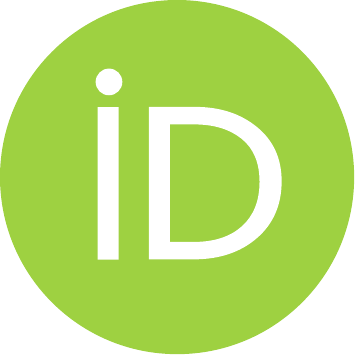}\hspace{1mm}Karin C. Knudson} \\
	Data Intensive Studies Center\\
	Tufts University\\
    Medford, MA 02155 \\
	\texttt{karin.knudson@tufts.edu} \\
	\And
	\href{https://orcid.org/0000-0002-8741-0621}{\includegraphics[scale=0.06]{orcid.pdf}\hspace{1mm}Anoopum S. Gupta} \\
	Department of Neurology\\
	Massachusetts General Hospital and Harvard Medical School\\
	Boston, MA 02114 \\
	\texttt{agupta@mgh.harvard.edu} \\\
}

\renewcommand{\shorttitle}{Assessing Cerebellar Disorders w/ IMU Data Using Time-Freq. and AR-HMM Approaches}

\hypersetup{
pdftitle={Assessing Cerebellar Disorders With Wearable Inertial Sensor Data Using Time-Frequency and AR-HMM Approaches},
pdfsubject={q-bio.NC, q-bio.QM, stat.AP, stat.ML},
pdfauthor={Karin C.~Knudson, Anoopum S.~Gupta},
pdfkeywords={ataxia, wearable sensors, IMUs, bayesian nonparametrics, time-frequency analysis, hidden Markov models},
}

\begin{document}
\maketitle

\begin{abstract}
	We use autoregressive hidden Markov models and a time-frequency approach to create meaningful quantitative descriptions of behavioral characteristics of cerebellar ataxias from wearable inertial sensor data gathered during movement. Wearable sensor data is relatively easily collected and provides direct measurements of movement that can be used to develop useful behavioral biomarkers. Sensitive and specific behavioral biomarkers for neurodegenerative diseases are critical to supporting early detection, drug development efforts, and targeted treatments. We create a flexible and descriptive set of features derived from accelerometer and gyroscope data collected from wearable sensors while participants perform clinical assessment tasks, and with them estimate disease status and severity. A short period of data collection ($<$ 5 minutes) yields enough information to effectively separate patients with ataxia from healthy controls with very high accuracy, to separate ataxia from other neurodegenerative diseases such as Parkinson's disease, and to give estimates of disease severity.
\end{abstract}


\section{Introduction}
Degenerative cerebellar ataxias are a heterogeneous group of disorders which include hereditary ataxias such as the spinocerebellar ataxias (SCAs), Friedrich's ataxia (FRDA), and ataxia-telangectasia (AT), as well as acquired ataxias such as idiopathic late-onset cerebellar ataxia (ILOCA) and multiple system atrophy (MSA). Cerebellar ataxias are relatively rare: AT affects 1-2.5 in 100,000, and the prevalence of the SCAs are estimated to range from 0 to 5.6 per 100,000 \citep{ruano2014global}. For comparison, Parkinson's disease, another neurodegenerative disease with movement-related hallmarks, is estimated conservatively at 0.3\% overall and $>3\%$ in those over 80 years old \citep{pringsheim2014prevalence}. Symptoms of ataxia include loss of balance and coordination, slurred speech, loss of fine motor skills, and difficulty walking. Symptoms and rate of progression of ataxia vary by individual and by type of ataxia. 
 
Sensitive, specific, and high resolution biomarkers for ataxia are necessary to support the development, deployment, and monitoring of (targeted) interventions, as well as for supporting early diagnosis efforts. The heterogeneity and rareness of cerebellar ataxias add to both the challenge and the importance of developing such biomarkers. 

Clinical scales are very important in the assessment of ataxia, and include SARA \citep{SchmitzHubsch2006Scale}, ICARS \citep{trouillas1997international}, and the Brief Ataxia Rating Scale (BARS) \citep{schmahmann2009development}.  Assessment via such scales requires a clinical visit, and progression from one discrete point to the next on the scale may happen slowly. Scores on these scales thus suffer from low resolution in two senses. First, the burden of a clinical visit means that there will generally be months between repeated assessments. Second, small but potentially meaningful degrees of progression in severity between visits may be missed due to scale coarseness.

There has been increasing interest in the use of wearable sensors, such as inertial measurement units (IMUs) that record accelerometer and gyroscope data, in assessing the progression of neurodegenerative diseases.  A goal of such research is to develop characterizations of movement-related symptoms in individuals that can serve as biomarkers and complement existing clinical scales in describing and quantifying disease features.  Data from IMUs give direct information about movement.  If such data could be used to provide fine-grained and nuanced descriptions of disease features and severity, they would allow for a clearer picture of a disease's natural history or response to treatment. 
Moreover, wearable sensor data from IMUs is gathered noninvasively and can be obtained not just in a clinical setting but also from patients at home; the relative ease of collection underscores the potential of wearable sensor based biomarkers to give a higher temporal resolution than what is possible from in-person clinical scoring. 

Hidden Markov models with hierarchical Dirichlet process priors (to accommodate an undetermined number of latent states) and added state ``stickiness" (to respect state persistence) are a Bayesian nonparametric approach building on hidden Markov models that can be used to segment sequential data (e.g. time series) and assign each segment to a latent state, the number of which need not be fixed beforehand \citep{fox2008hdp}. Sticky HDP-HMMs have been extended to include switching dynamics that govern the observed data points via autoregressive (AR) processes \citep{fox2009nonparametric}. The resulting AR-HMMs would seem well-suited to aid in the generating rich and flexible representations of movement data as they can be used to simultaneously learn a segmentation of movements into short submovements, the number of submovements, and the dynamics of the submovements. We explore their utility in this context in the current work.

Our contributions can be summarized as follows:
\begin{itemize}
    \item We demonstrate that an autoregressive hidden Markov model (AR-HMM) that allows for nonparametric extension can be used to characterize submovements within clinical motor assessments recorded by IMUs. To our knowledge, this paper represents the first application of such models to human motion measured by IMUs.
    \item We present two approaches to developing meaningful and descriptive quantifications of movement tasks recorded by IMUs, one set learned from data based on an autoregressive hidden Markov model, and another based on a time-frequency approach.
    \item We apply these approaches to clinical data and demonstrate excellent classification accuracy and good severity score estimation accuracy for ataxia when the AR-HMM and time-frequency features are combined and used with classical machine learning approaches (random forests).
\end{itemize}

\section{Related Work}

As the technologies for wearable inertial sensors such as gyroscopes and accelerometer have become more advanced and widespread in recent years, research into their potential for characterizing and quantifying neurodegenerative diseases has increased.  A growing volume of work has emerged around the potential of inertial wearable sensors in quantifying Parkinson's disease in particular; for reviews, see \cite{rovini2017wearable, maetzler2013quantitative}.  Inertial sensors have been used for gait analysis \citep{klucken2013unbiased}, and the identification of tremor \citep{barrantes2017differential} and bradykinesia \citep{lonini2018wearable}.  Inertial sensor data has shown potential in both identifying individuals with Parkinson's \citep{butt2017biomechanical} and quantifying its severity \citep{zhan2018using}.  Studies have explored data gathering not just in the clinic, but in at-home settings as well \citep{battista2018novel, heijmans2019evaluation}.  

The literature on inertial sensor data for quantifying ataxia (to which the present work contributes) is a notably smaller body of work, with
the greatest area of focus so far on measuring features of gait \citep{hickey2016validity, lemoyne2016wearable, phan2019random, phan2019quantitative, ilg2020real}.  IMU measurements of finger rhythmic tapping with the finger and/or foot have also been used to predict the presence and severity of ataxia \citep{tran2019automated, nguyen2018quantitative}. In the gait and finger tapping settings, orientation of the sensors relative to the surroundings stays relatively constant - more so than in the clinical tasks we consider in this work. The features used to train models to predict ataxia status and severity based on IMU data have included features based on multiscale entropy or fuzzy entropy \citep{tran2019automated, nguyen2018quantitative, phan2019quantitative}, inter-movement intervals \citep{tran2019automated, gavriel2015towards}, and features calculated from the frequency domain \citep{nguyen2018quantitative, phan2019random}. In Parkinson's disease, CNN-based approaches have also been used \citep{lonini2018wearable}.

Initially applied to segment audio files for speaker diarization \citep{fox2008hdp}, sticky HDP-HMMs have also been used in chunking human motion data from a motion capture system for imitation learning \citep{taniguchi2011unsupervised}. Their extensions to include dynamics governed by autoregressive processes \citep{fox2009nonparametric} have been used to characterize mouse motion (captured with video) as a combination of discrete movement modules \citep{wiltschko2015mapping} in ways shown to be informative e.g. about drug identity, dose, and class when diverse mouse behavior was generated via pharmacology \citep{wiltschko2020revealing}.

\section{Data collection and participants}
Participants wore six APDM sensors: one sensor on each wrist and ankle, a sensor in the pocket, and a lumbar sensor.  The wearable sensors captured data at a sampling frequency of 128 Hz along three axes from an accelerometer, gyroscope, and magnetometer ---for a total of nine channels per sensor--- as participants performed a variety of clinical tasks. The analysis in this work uses accelerometer and gyroscope data from the two wrist sensors during the finger-nose-finger task (reaching to touch a tablet computer) and light bulb tasks, and from two ankle sensors during the heel-shin task. These tasks were selected from among those performed because they together represented upper and lower extremity function and because the motions associated with them are rhythmic and repetitive, and thus our choices of frequency-based characteristics were naturally applicable.
Data was collected from 212 participants and 264 sessions (with repeat sessions for some subjects, typically recorded months apart), including control subjects and those with ataxia and parkinsonism. For the analysis, we use only subjects from which we have recordings for for all three tasks: finger nose, light bulb, and heel shin (195 subjects, 242 sessions). See Table \ref{tab:demograhics} for demographic information about the subjects.

\begin{table*}
\caption{Demographic characteristics of subjects. Specific ataxia diagnoses are shown in separate lines for each diagnosis represented by more than one subject. Note that mean severity score for parkinsonism is reported on the UPDRS \cite{goetz2008movement}; ataxia severity scores are on the BARS scale.}
	\centering
\label{tab:demograhics}
\begin{tabular}{  l  l c c c c } 
 \toprule & n & Age & Women & Men & Severity score \\ 
& subjects, sessions & mean (SD) &  &  & mean (SD) \\ 
\midrule
Total &  195, 242 & 46.9 (24.1) & 78 & 117 & n/a\\ 
 \toprule
Ataxia (all) &  109, 144 & 43.1 (23.8) & 49 & 60 & 10.3 (5.3)\\
AT &  34, 56 & 12.4 (6.1) & 13 & 21 & 11.3 (5.7)\\
Episodic Ataxia &  2, 2 & 50.0 (28.0) & 0 & 2 & 1.2 (0.2)\\
FA &  2, 2 & 55.5 (5.5) & 1 & 1 & 16.2 (2.8)\\
MSA &  5, 6 & 60.4 (5.2) & 2 & 3 & 12.2 (3.8)\\
SCA 1 &  4, 8 & 49.0 (14.1) & 3 & 1 & 8.1 (2.0)\\
SCA 2 &  2, 2 & 61.0 (6.0) & 1 & 1 & 12.0 (1.5)\\
SCA 3 &  11, 14 & 51.0 (10.6) & 9 & 2 & 9.5 (3.3)\\
SCA 6 &  8, 9 & 69.0 (6.0) & 3 & 5 & 12.8 (6.1)\\
SPG 7 &  2, 2 & 58.0 (2.0) & 0 & 2 & 9.0 (1.0)\\
Transient Ataxia &  2, 2 & 72.5 (2.5) & 1 & 1 & 1.0 (0.0)\\
Ataxia (other) &  36, 40 & 55.8 (14.0) & 16 & 20 & 10.6 (5.7)\\ 
 \toprule
Parkisonism &  52, 59 & 67.6 (7.9) & 14 & 38 & 16.5 (9.7)\\ 
 \toprule
Control &  34, 39 & 27.5 (18.4) & 15 & 19 & n/a\\ 
 \bottomrule
\end{tabular}
\end{table*}

\section{Methods}

\subsection{Preprocessing}
For each individual's performance of a task, principal component analysis was used on the three dimensional accelerometer and gyroscope signals (separately) from each sensor, and the loadings for the first principal component were retained.  This operation reduced the dimensionality of the signal threefold at each time point and also removed the dependence of the signal on sensor orientation.  Signals were divided temporally into equal `rest' and `task' portions based on the total magnitudes of velocity in the signal, so that, for example, a sensor on the right lower extremity during the heel-shin task would include the `task' portion where the task was performed with the right leg, and the `rest' portion when the task was being performed with the left leg.  This division was made according to the above criterion for each individual and task to account for occasional inconsistencies where a subject started their motion on the opposite side from what was instructed.

For the AR-HMM based features, we additionally included a wavelet denoising step with a Symlets-4 wavelet and a threshold of 0.04 on the vector of loadings of the first principal component \citep{lee2019pywavelets} and, to speed computation in the statistical learning of the AR-HMM we downsampled each time series by a factor of ten.

\subsection{Time frequency features}

\begin{figure}
  \centering
  \includegraphics[width=0.6\linewidth]{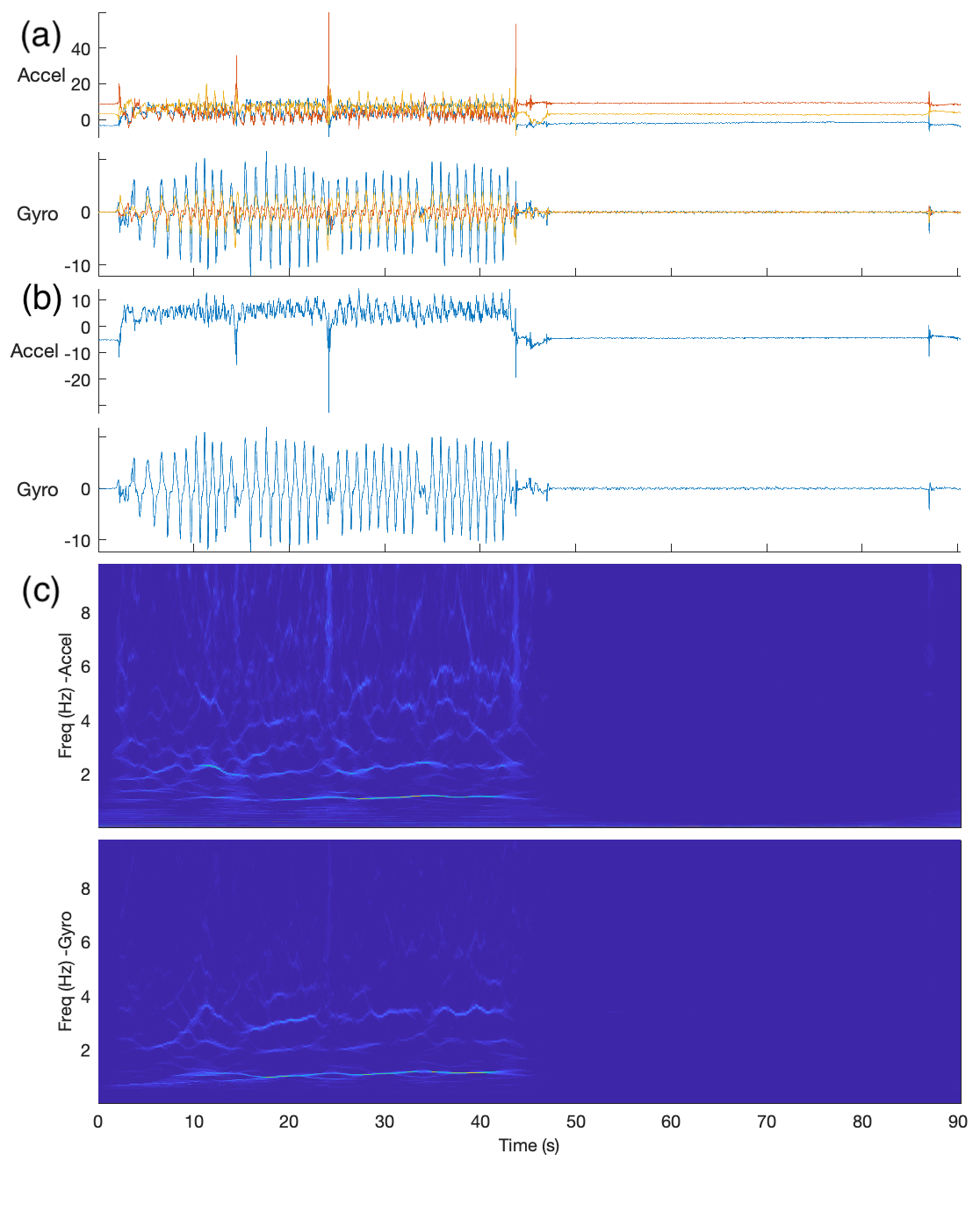}
  \caption{(a): Raw IMU signals for a control subject performing the finger nose finger task with their right hand. (b): Signals projected along first principal component. (c): Magnitude of synchrosqueezed transform (SST) of projected signals. Note: During the right half of the graphs, the measured hand was at rest. On the left hand side, a noticeable feature is the horizontal lines around 1Hz, the approximate frequency with which the subject is completing the repeated motion.}
\label{fig:sst}
\end{figure}

For the time-frequency based features, a wavelet synchrosqueezed transform \citep{daubechies2011synchrosqueezed} with an analytic bump wavelet was applied to a one dimensional data vector for each session and task, sensor position (right or left relevant extremity), and modality (accelerometer or gyroscope) combination to yield a time-frequency distribution of power for each signal (Figure \ref{fig:sst})  \citep{matlabwavelet}. Compared to a wavelet transform, the wavelet synchrosqueezed transform includes a ``reallocation" along the frequency access that tends to sharpen the result in the frequency direction. In the context of rhythmic tasks and neurodegenerative diseases that can affect the rhythmicity and speed of actions, we deemed sharpness in the frequency direction desirable. Frequencies above 15 Hz were omitted from the analysis.

The following 16 features were calculated for each combination of task (N=3), sensor position (N=2, left and right side), and sensor modality (N=2, accelerometer and gyroscope) for a total of 192 features.  Note that some features are calculated using frequency thresholds: above or below 2Hz for finger nose and heel shin, and 3Hz for light bulb.  The cutoffs were chosen because the frequencies of the tasks themselves (e.g. rate of reaching in the finger-nose task and rotating in the light-bulb task) were deemed to be lower than 2 or 3 Hz respectively, while frequencies of a tremor would generally fall above this cutoff.
\begin{itemize}
    \item Total power (rest and task): Power summed across all time bins and all frequencies.
    \item Ratio of task to rest power: Total power during task divided by total power during rest.
    \item Low frequency power (rest and task): Power summed across all time bins and all frequencies below the cutoff (2 or 3 Hz, depending on task).  
    \item High frequency power (rest and task):  Power summed across all time bins and all frequencies above cutoff.  
    \item Ratio of low to high frequency power (rest and task): Low frequency power divided by high frequency power. 
    \item Center frequency (rest and task): Weighted sum of frequency (with weights the total power for that frequency summed over time), divided by summed power over all frequency.
    \item Spread of frequency (rest and task): Squared difference of frequency from center frequency, multiplied total power for that frequency and then summed over all frequencies. 
    \item Center frequency of low frequency (task): As above, but calculated only for frequencies below cutoff. 
    \item Center frequency of high frequency (task): As above, but calculated only for frequencies above cutoff.
    \item Cosine similarity of adjacent time bins (task): Mean cosine similarity between the vectors of powers for each frequency for adjacent time bins.
\end{itemize}

These features were designed before the downstream analysis was performed, and were chosen to capture known features of ataxia, such as changes in the rhythmicity of movement.

\subsection{Autoregressive Hidden Markov model}
To construct a second, complementary set of features, we turned to a different paradigm.  

\begin{figure}[ht]
  \centering
  \includegraphics[width=0.5\linewidth]{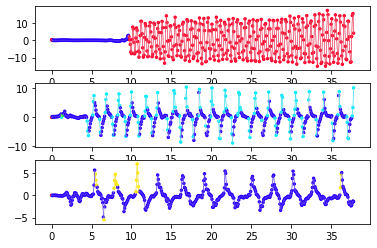}
  \caption{An example time series in which each time point in a preprocessed gyroscope signal has been assigned a learned latent state (state represented by color). Top: light bulb task (subject with mild ataxia). Middle: finger nose (control subject). Bottom: finger nose (subject with ataxia).}
\label{fig:example_coded_states}
\end{figure}

In the AR-HMM model, each time point is assumed to have an unobserved underlying state that evolves as a Markov chain subject to unobserved probabilities of transitions between states at each point.  Given the discrete state associated with a time point, the observed data (accelerometric and gyroscopic) is assumed to proceed via an autoregressive process with parameters that are characteristic to that state. Thus, broadly speaking, when we perform inference to learn the sequence of states for motion data of a particular individual and task, each learned state is a label related to the characteristics of the motion in the preceding time bins (Figure \ref{fig:example_coded_states}). The Markov model that governs transitions between states here includes a stickiness parameter \citep{fox2008hdp}, which boosts the number of self-transitions so that states tend to persist over time. Importantly, the states and their associated dynamics are learned from the data, and the state learning can proceed non-parametrically, so that the number of states needed to effectively describe the data is also learned from the data. Applying this model to motion data thus gives us a picture of the motion as a sequence of submovements, with each submovement corresponding to a period of time where the latent state is constant and the motion evolves following the particular dynamics associated with that state. 

Concretely, the model for the states $x_t$ and the observed multidimensional (here 2 dimensional, after pre-processing) IMU data vectors $\vec{y}_t$, both indexed by time, is given by:
$$ p(x_t | x_{t-1}, \mathbf{\pi}) = \pi_{x_t, x_{t-1}} $$
$$ \vec{y}_t| \mathbf{A}^{(x_t)}, \mathbf{\Sigma}^{(x_t)}, \mathbf{y}_{t-1}, x_t \sim$$ $$ \mathrm{MVN}(\mathbf{A}^{(x_t)} \mathbf{Y}_{t-1:t-n}, \Sigma^{(x_t)} )$$
 
 Here $\pi$ is the $L \times L$ transition matrix, where $L$ is the number of discrete states and $x_t \in \{1,...,L\}$. Learned parameters $\mathbf{A}^{(x_t)}, \mathbf{\Sigma}^{(x_t)}$ come from a set of $L$ pairs $$(\mathbf{A}^{(1)}, \mathbf{\Sigma}^{(1)}),...(\mathbf{A}^{(L)}, \mathbf{\Sigma}^{(L)})$$ that govern the dynamics. MVN denotes the multivariate normal distribution.  By $\mathbf{Y}_{t-1:t-n}$ we denote the observed data vectors from the preceding $n$ time points. For the present application, we set the number of time lags in the model to  $n=5$, so that the next time point evolves in accordance with the previous 0.4 seconds of data. We also explored $n=4$ and $n=6$ lags, with results that appeared qualitatively similar. Across a broader range, however, we expect the choice of number of lags to have a strong effect on the model and on the choice of other parameters (for example, adding more lags allows additional complexity in the dynamics, suggesting a corresponding increase in the number of latent states might be needed in order to capture this complexity).
 
 We are following a procedures common in other applied uses of HDP-HMMs by  truncating to a finite number of states $L$ and using Dirichlet distribution priors \citep{fox2008hdp, fox2009nonparametric, wiltschko2015mapping}. Note that as $L \to \infty$, the finite hierarchical model converges in distribution to a hierarchical Dirichlet process prior \citep{teh2005sharing}.
 
 Let $\kappa$ be a stickiness parameter that will be chosen empirically (larger $\kappa$ will encourage runs time points in the same state), and let $\vec{\delta}_i$ be the unit vector such that $\delta_{i,j} = 1$ when $i=j$ and 0 otherwise. We assume the following prior distributions for $\pi, \mathbf{A}, \mathbf{\Sigma}$:
 $$ \mathbf{A}, \mathbf{\Sigma} \sim \mathrm{MNIW}(\nu_0, \mathbf{S}_0, \mathbf{M}_0, \mathbf{K}_0)$$
 $$ \vec{\beta}_i \sim \mathrm{Dirichlet}(\gamma/L, ... \gamma/L ) $$
 $$ \vec{\pi}_{i \cdot} \sim \mathrm{Dirichlet}(\alpha \vec{\beta} + \kappa \vec{\delta}_i) $$
MNIW denotes the matrix normal inverse Wishart distribution.
We set hyperparameters $\nu_0$, $\mathbf{S}_0$, $\mathbf{M}_0$, $\mathbf{K}_0$, $\alpha$, $\gamma$, $\kappa$, as follows, although these hyperparameters could be given their own prior distributions if desired. 
We let $\alpha=\gamma=\kappa=20$, based on prior predictive sampling that gave frequent runs of states (interpreted as submovement length) that were centered around 100-200ms in length (median run length of 156ms on a 30s simulated interval). We set $L=5$, because with these hyperparameter values and larger number of states (6 or 8), over $98\%$ of states in the training data were assigned to one of the five most common states. Hyperparameters could be changed to encourage more states, or the model could be extended to allow for an indeterminate number of states. In our current setting, we were content with a small number of states whose associated dynamics could be more easily inspected. We set the hyperparameters for the autoregressive parameters to: 
$\nu_0 = 5, \, \mathbf{S}_0 = 0.01 \mathbf{I}, \mathbf{M}_0=.25  \mathbf{1}_{2 \times 2n}$ (where  $\mathbf{I}$ denotes the identity matrix and $\mathbf{1}_{2 \times 2n}$ denotes the $n \times 2n$ matrix of all ones). We let $K_0$ be a diagonal matrix whose entries are linearly spaced between 5 and 100.

\begin{figure}
  \centering
    \includegraphics[width=0.6\linewidth]{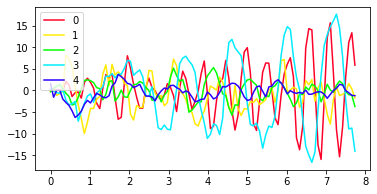}
  \caption{Sample trajectories evolving according to the estimated AR parameters $(\vec{A}^{(0)}, \vec{\Sigma}^{(0)}),...(\vec{A}^{(4)}, \vec{\Sigma}^{(4)})$ for each of the five states. Visual inspection of such trajectories are a simple tool to help add insight about the types of movement corresponding to each state.}
\label{fig:sample_trajectories}
\end{figure}

Inference for the posterior distributions proceeded via Gibbs sampling, as in \citep{fox2009nonparametric, wiltschko2015mapping}. However, to reduce computation time, the parameters $\mathbf{A}, \mathbf{\Sigma}$ that controlled the autoregressive process were first learned (via Gibbs sampling) from a subset of the data that was chosen to represent the range of common submovements. This subset of data included a recording from each of the three tasks for 4 subjects: a control, and three ataxic subjects with BARS scores ranging from 1 to 20.
We sampled to obtain 2000 draws, and used a burn-in of 100 samples. Mode switching was a concern, since there are symmetries in the labeling of the states, and the posterior means of the autoregressive parameters would feed into the next step of sampling. Visual inspection of the sampled $\beta$ parameters showed an obvious mode switch around 1000 samples in, so we truncated to the first 1000 samples (with a burn-in of 100 samples) in computing the posterior means of the autoregressive parameters.
See Figure \ref{fig:sample_trajectories} for sample trajectories evolving according to the the learned dynamics associate with each state, as measured by their posterior means.

We then proceeded with the posterior means of  $(\mathbf{A}^{(1)}, \mathbf{\Sigma}^{(1)}),..., (\mathbf{A}^{(L)}), \mathbf{\Sigma}^{(L)})$ as fixed autoregressive parameters (and using the posterior means for $\beta$ and $\pi$ from the first stage of sampling as initial values in the Gibbs sampler) and learned the state sequences $x_t$, the transition probabilities, $\vec{\pi}$, and parameters $\vec{\beta}$ for each individual, task, and sensor. In this second stage of Gibbs sampling, for the state-related parameters, we drew 200 samples, with a burn-in of 50 samples.

\vspace{1cm}
The following features were calculated from each sequence of state sample modes: 
\begin{itemize}
\item the frequencies of each state in the data 
\item each state's estimated self-transition probability
\item the mean of the the length of runs (consecutive appearances) of each state
\item the standard deviation of the length of runs (consecutive appearances) of each state
\item the estimated entropy rate of the state sequence as calculated from the estimated Markov transition probabilities. 
\end{itemize}

We also use both the state samples and their modes to calculate how concentrated on the mode the drawn state samples were:
\begin{itemize}
\item for each state, averaged over all time points for which that state was the mode, the proportion of state samples drawn that were equal to that mode
\item the proportion of all state samples drawn that were equal to the corresponding-in-time mode state
\end{itemize}
We thus calculate 5 state-specific features (for each of L=5 states) and 2 other features for 27 features from each of N=3 tasks and N=2 sensors (left and right), yielding a total of 162 AR-HMM derived features.

\begin{figure}
  \centering
  \includegraphics[width=0.6\linewidth]{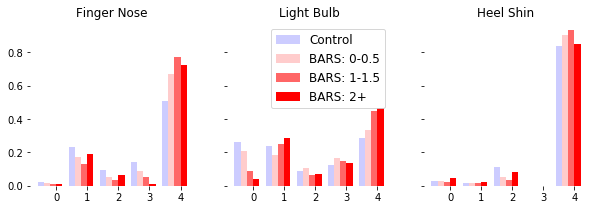}
  \caption{Frequency of occurrences of the latent states (numbered 0-4) present in sessions corresponding to different tasks and BARS subscores (BARS right arm subscore for finger nose and light bulb, BARS right leg subscore for heel shin). The dynamics associated with a state are common across all tasks. Note the variation in frequency of state based on task and severity of ataxia score. For instance, within the light bulb task there seems to be a systematic trade off between the frequency of occurrence of states 0 and 4 as the BARS arm score increases.}
\label{fig:state_frequencies}
\end{figure}

Many other derived features are possible; we chose a naive set in order to form a baseline and demonstrate the broad applicability of the AR-HMM-based coding of states to the disease status and severity. As a visual check of the relevance of the coded states, sample trajectories for each state were visualized (Figure \ref{fig:sample_trajectories}). Note that the frequency of each state varied by task and diagnosis (Figure \ref{fig:state_frequencies}).

\subsection{Cross validation}
All results are computed and reported based on a leave-one-subject-out cross-validation setup, performing model parameter selection and model training with all session data from one subject held out, using the selected model to predict the disease status or severity score for the session(s) for that one subject, and then aggregating measures of the correctness of the predictions across all the predictions for held-out subjects in order to report the relevant accuracy metrics.
 
\subsection{Classification}

Classification of disease status, such as ataxia vs. control or ataxia vs. parkinsonism, is performed using a balanced random forest with 200 trees.  The balanced random forest modifies a random forest with undersampling to compensate for the imbalanced classes; in our case, the imbalance is that there were more recordings from ataxic subjects than from controls \citep{chen2004using, imblearn}. 

\subsection{Score Prediction}
\label{sec:score_pred}
BARS scores were estimated by training a random forest model with 200 trees, a maximum tree depth of 10, and a mean absolute error (MAE) criterion \citep{scikit-learn}.  Predictions were performed for BARS total score (obtained for 133 out of 183 sessions), and two motor subscores: the BARS right arm score (obtained for 155 out of 183 sessions) and, because of a lack of gait scoring for some subjects, an arm-and-leg subscore consisting of the sum of the BARS arm and leg score (obtained for 142 out of 183 sessions). Total BARS scores (which sum oculomotor and speech subscores with clinical scores related to arms, legs, and gait) range from 0 to 30, with higher numbers indicated greater severity. The BARS right arm subscore ranges from 0 to 4, while the sum of the arm and leg subscores range from 0 to 16.

\section{Results}
\subsection{Classification Results}
\begin{figure}
  \centering
  \includegraphics[width=0.21\linewidth]{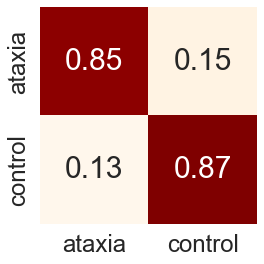}
 \includegraphics[width=0.21\linewidth]{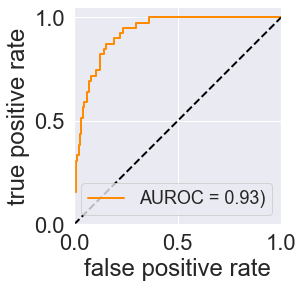}
  \includegraphics[width=0.21\linewidth]{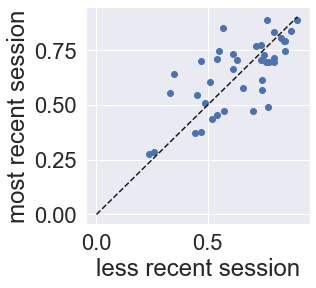}
 \includegraphics[width=0.21\linewidth]{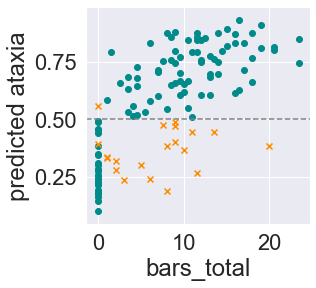}
  \caption{Classification performance distinguishing ataxic subjects from control subjects.  From left to right: normalized confusion matrix (rows give true classification, columns give prediction); ROC curve; Test-retest correlation of classification scores, where scores closer to 1 represent a prediction of ataxia and scores closer to 0 prediction of control; Total BARS score vs. predicted probability of ataxia ($>0.5$ corresponds to a prediction of ataxia, X = incorrect prediction).}
\label{fig:classification_acc}
\end{figure}

Classification results were highly accurate, with classification of ataxia against controls achieving an area under the ROC curve (AUROC) of 0.93, a sensitivity of 0.85 and specificity of 0.87. (Figure \ref{fig:classification_acc}). The test-retest correlation of classification scores calculated from the 35 sessions of subjects who had repeated visits was 0.72.

The relative importance of the time-frequency and AR-HMM based features was evaluated. Out of the top 20 features by importance in the classifier (ataxia vs. control), 11 features come from the time-frequency approach, and 9 come from the AR-HMM approach. 13 of these top 20 features came from the finger nose task, 7 from the light bulb task, and 0 from the heel shin task. When classification was performed using only features from the time-frequency approach, the AUROC for the ataxia vs. control classification task was 0.93 (sensitivity: 0.86, specificity: 0.90).  When classification was performed using only freatures derived from the AR-HMM approach, the AUROC for the ataxia vs. control classification task was 0.89 (sensitivity: 0.81, specificity: 0.80).

Since one potential use for wearable-sensor based measurement of disease is early detection, we also considered the results of training a classifier to detect ataxia when set of subjects is limited to mild cases (here we take mild to mean having a total BARS score of less than 10). For this task, the AUROC score was 0.87 (sensitivity: 0.76, specificity: 0.82). 
We also consider separately the classification tasks for adult ataxia vs. control and pediatric ataxia vs. control. Here we define subjects under age 18 to be pediatric. Performance  of the classifier was better when limited to pediatric subjects, most of whom have ataxia-telangiectasia (AT) and more severe ataxia. 

\begin{table*}
  \caption{Metrics for binary classification of general diagnosis}
  \label{tab:freq}
  \centering
  \begin{tabular}{lcccc}
    \toprule
    & AUROC & Sensitivity & Specificity & Test-retest correlation (\# sessions)\\
    \midrule
    Ataxia/control & 0.93 & 0.85 & 0.87 &  0.70 (35)\\
    Mild ataxia/control & 0.87 & 0.76  & 0.82 & 0.62 (12) \\
    Pediatric ataxia/control& 0.95  & 0.90 & 0.93 & 0.84 (20)\\
    Adult ataxia/control & 0.90 & 0.80 & 0.80 & 0.46 (15)\\
    Ataxia/control (AR-HMM only)& 0.89  & 0.81 & 0.80 & 0.51 (35) \\
    Ataxia/control (time-freq only)& 0.93 & 0.86 & 0.90 & 0.74 (35) \\
    Ataxia/parkinsonism & 0.93 &  0.88 & 0.83 & 0.94 (35) \\
  \bottomrule
\end{tabular}
\end{table*}

\begin{figure}
  \centering
 \includegraphics[width=0.21\linewidth]{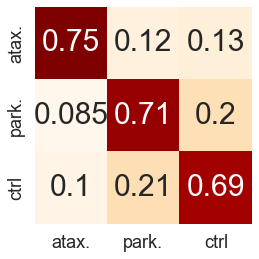}
   \includegraphics[width=0.21\linewidth]{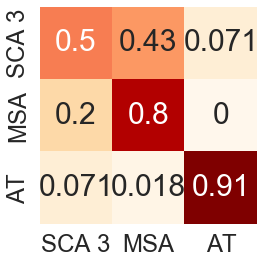}
  \caption{Left: normalized confusion matrix for classification performance distinguishing ataxia ataxia, parkinsonism, and control. Right: classification of three specific ataxia diagnoses. Rows are true classification, columns are predicted classification.}
\label{fig:class_ataxia_parkinsonism_and_subtypes}
\end{figure}

Classification of ataxia vs. parkinsonism was also highly accurate (AUROC 0.93: sensitivity 0.88: specificity: 0.83) suggesting that the features described above do more than just demonstrate an overall motor impairment, but can describe the type of impairment. In the ternary classification ataxia vs. parkinsonism vs. control subjects, we observed an accuracy of rate of 0.70 (Figure \ref{fig:class_ataxia_parkinsonism_and_subtypes}).

We also assess the applicability of the features presented here to distinguishing between different types of ataxia. In the task of classifying ataxia-telangiectasia (AT), multiple system atrophy (MSA) and spinocerebellar ataxia type 3 (SCA 3, the most represented of the SCAs in our data), we obtain an accuracy of 0.80. (Figure \ref{fig:class_ataxia_parkinsonism_and_subtypes}).

\subsection{Score Prediction Results}

We predict BARS total score and subscores as described in section \ref{sec:score_pred}, trained on the combined set of all control subjects (who have BARS scores of 0) and all subjects with ataxia.

\begin{figure}[ht]
   \centering
   \includegraphics[width=0.21\linewidth]{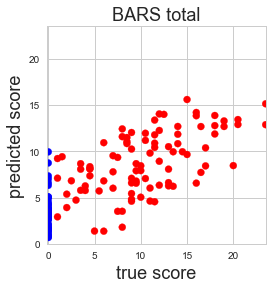}
  \includegraphics[width=0.21\linewidth]{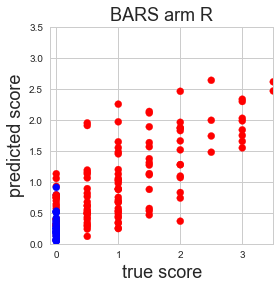}
  \includegraphics[width=0.21\linewidth]{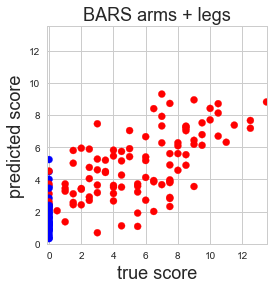}
   \includegraphics[width=0.21\linewidth]{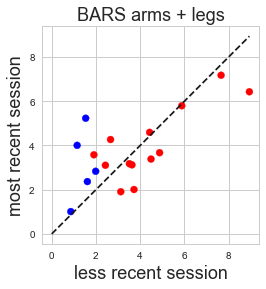}
   \caption{From left to right: true vs. predicted total BARS scores; true vs. predicted BARS right arm subscore; rue vs. predicted BARS arms and legs subscore; test-retest reliability of BARS arms and legs subscores for subjects with multiple visits. (Red=ataxia, blue=control).}
 \label{fig:score_prediction}
 \end{figure}

The correlation of predicted and true total BARS score was 0.74 ($R^2 = 0.53$)  with a mean absolute error (MAE) of 3.76 points and a test-retest correlation 0.72 from 16 subjects with multiple visits),. The correlation of predicted and true BARS motor arm and leg subscore was 0.75 ($R^2 = 0.54$, MAE = 2.14, test-retest correlation 0.70 from 16 subjects with multiple visits). The correlation of predicted and true right arm BARS score is 0.77 ($R^2 = 0.59$, MAE = 0.47, test-retest correlation 0.96 from 20 subjects with multiple visits) (Figure \ref{fig:score_prediction}). 

\section{Discussion}
Our results demonstrate the relevance of two particular approaches for representing movements measured by IMUs in the context of describing neurodegenerative diseases, particularly cerebellar ataxias. 

The two approaches involve performing a synchrosqueezed wavelet transform and training an AR-HMM. We find that the features derived from each approach are highly informative about an individual's diagnosis, even in the context of short data recordings (under 5 minutes per participant). In addition to allowing for high rates of success classifying ataxic subjects against healthy controls, models trained on these features were able to perform several more subtle tasks: classifying mild ataxia and healthy controls, classifying ataxia and parkinsonism, and distinguishing several types of ataxia. Additionally, we found that regression models trained on AR-HMM and time-frequency features had good performance in estimating disease severity as measured by neurologist-administered clinical scales.

The current work points to the potential of these two classes of features derived from wearable sensor data to support important applications: as biomarkers for understanding an individual's disease and its progression and to support targeted interventions, effective clinical trials, and early diagnosis.

We used IMU measurements from three different clinical tasks. Our results suggested that the light bulb and finger nose clinical tasks were more informative than the heel shin task for classification and severity prediction when using the chosen feature sets. Additional work with multiple clinical tasks and other kinds of movement will be important in determining which types of movements and tasks can be used to most efficiently gather information about  pertinent behavioral aspects of an individual's neurodegenerative disease.

The time-frequency features derived from the synchrosqueezed transform were used to train effective classifiers on their own and in combination with the AR-HMM-based features. Our time frequency approach is not far from baseline techniques; at least one other work has used wavelet transforms \citep{oung2018empirical}, and more have used methods from the general category of time-frequency approaches in the setting of gaining insight into neurodegenerative diseases via wearable sensor measurements. Our use of the synchrosqueezed transform as a starting point and particular choice of the derived feature set is novel, chosen with an eye towards simplicity and interpretability; we hope they can add to the body of understanding of features relevant to this this domain. 

Using submovements learned with an AR-HMM was novel in the setting of human IMU data and neurodegenerative diseases, and our results point to their applicability in this domain. An exciting aspect of AR-HMMs in this realm of application is their extensibility and flexibility. AR-HMMs can be extended nonparametrically to learn an appropriate number of states, and thus logically extend to encompass more kinds of movement. Moreover, the AR-HMM features are task-agnostic, facilitating the future incorporation of additional behavioral tasks and even free behavior. ?
Future work with AR-HMMs in the context of IMU data and neurodegenerative diseases could involve deeper investigation into the choices of hyperparameters or hyperpriors that are appropriate in the setting of IMU data from a range of clinical tasks, as well as for capturing more varieties of submovements within broader sets of tasks or potentially free-living behavior.

 Both the synchrosqueezed transform and the AR-HMM can serve as starting points for a rich set of features that go beyond what were considered here. We hope that future work could consider our two approaches as preprocessing steps, and explore the most relevant derived features in different settings and for different purposes related to understanding and supporting effective interventions for cerebellar ataxias and other neurodegenerative diseases.

\section*{Acknowledgements}
We would like to thank Mary Donovan, Nergis Khan, and Winnie Ching for their help collecting data, Pavan Vaswani for designing the finger-nose-finger stimuli, and Jeremy Schmahmann, Albert Hung, and Christopher Stephen for their help in recruiting research participants and generating clinical scores. The study was generously supported by the Ataxia-Telangiectasia Children's Project and Biogen.

\bibliographystyle{unsrtnat}
\bibliography{references} 

\begin{thebibliography}{36}
\providecommand{\natexlab}[1]{#1}
\providecommand{\url}[1]{\texttt{#1}}
\expandafter\ifx\csname urlstyle\endcsname\relax
  \providecommand{\doi}[1]{doi: #1}\else
  \providecommand{\doi}{doi: \begingroup \urlstyle{rm}\Url}\fi

\bibitem[Ruano et~al.(2014)Ruano, Melo, Silva, and Coutinho]{ruano2014global}
Luis Ruano, Claudia Melo, M~Carolina Silva, and Paula Coutinho.
\newblock The global epidemiology of hereditary ataxia and spastic paraplegia:
  a systematic review of prevalence studies.
\newblock \emph{Neuroepidemiology}, 42\penalty0 (3):\penalty0 174--183, 2014.

\bibitem[Pringsheim et~al.(2014)Pringsheim, Jette, Frolkis, and
  Steeves]{pringsheim2014prevalence}
Tamara Pringsheim, Nathalie Jette, Alexandra Frolkis, and Thomas~DL Steeves.
\newblock The prevalence of parkinson's disease: A systematic review and
  meta-analysis.
\newblock \emph{Movement disorders}, 29\penalty0 (13):\penalty0 1583--1590,
  2014.

\bibitem[Schmitz-H{\"u}bsch et~al.(2006)Schmitz-H{\"u}bsch, du~Montcel, Baliko,
  Berciano, Boesch, Depondt, Giunti, Globas, Infante, Kang, Kremer, Mariotti,
  Melegh, Pandolfo, Rakowicz, Ribai, Rola, Sch{\"o}ls, Szymanski, van~de
  Warrenburg, D{\"u}rr, and Klockgether]{SchmitzHubsch2006Scale}
T.~Schmitz-H{\"u}bsch, S.~Tezenas du~Montcel, L.~Baliko, J.~Berciano,
  S.~Boesch, C.~Depondt, P.~Giunti, C.~Globas, J.~Infante, J.~S. Kang,
  B.~Kremer, C.~Mariotti, B.~Melegh, M.~Pandolfo, M.~Rakowicz, P.~Ribai,
  R.~Rola, L.~Sch{\"o}ls, S.~Szymanski, B.~P. van~de Warrenburg, A.~D{\"u}rr,
  and T.~Klockgether.
\newblock Scale for the assessment and rating of ataxia.
\newblock \emph{Neurology}, 66\penalty0 (11):\penalty0 1717--1720, 2006.

\bibitem[Trouillas et~al.(1997)Trouillas, Takayanagi, Hallett, Currier,
  Subramony, Wessel, Bryer, Diener, Massaquoi, Gomez,
  et~al.]{trouillas1997international}
P~Trouillas, T~Takayanagi, M~Hallett, RD~Currier, SH~Subramony, K~Wessel,
  A~Bryer, HC~Diener, S~Massaquoi, CM~Gomez, et~al.
\newblock International cooperative ataxia rating scale for pharmacological
  assessment of the cerebellar syndrome.
\newblock \emph{Journal of the neurological sciences}, 145\penalty0
  (2):\penalty0 205--211, 1997.

\bibitem[Schmahmann et~al.(2009)Schmahmann, Gardner, MacMore, and
  Vangel]{schmahmann2009development}
Jeremy~D Schmahmann, Raquel Gardner, Jason MacMore, and Mark~G Vangel.
\newblock Development of a brief ataxia rating scale (bars) based on a modified
  form of the icars.
\newblock \emph{Movement disorders}, 24\penalty0 (12):\penalty0 1820--1828,
  2009.

\bibitem[Fox et~al.(2008)Fox, Sudderth, Jordan, and Willsky]{fox2008hdp}
Emily~B Fox, Erik~B Sudderth, Michael~I Jordan, and Alan~S Willsky.
\newblock An hdp-hmm for systems with state persistence.
\newblock In \emph{Proceedings of the 25th international conference on Machine
  learning}, pages 312--319, 2008.

\bibitem[Fox et~al.(2009)Fox, Sudderth, Jordan, and
  Willsky]{fox2009nonparametric}
Emily Fox, Erik~B Sudderth, Michael~I Jordan, and Alan~S Willsky.
\newblock Nonparametric bayesian learning of switching linear dynamical
  systems.
\newblock In \emph{Advances in neural information processing systems}, pages
  457--464, 2009.

\bibitem[Rovini et~al.(2017)Rovini, Maremmani, and Cavallo]{rovini2017wearable}
Erika Rovini, Carlo Maremmani, and Filippo Cavallo.
\newblock How wearable sensors can support parkinson's disease diagnosis and
  treatment: a systematic review.
\newblock \emph{Frontiers in neuroscience}, 11:\penalty0 555, 2017.

\bibitem[Maetzler et~al.(2013)Maetzler, Domingos, Srulijes, Ferreira, and
  Bloem]{maetzler2013quantitative}
Walter Maetzler, Josefa Domingos, Karin Srulijes, Joaquim~J Ferreira, and
  Bastiaan~R Bloem.
\newblock Quantitative wearable sensors for objective assessment of parkinson's
  disease.
\newblock \emph{Movement Disorders}, 28\penalty0 (12):\penalty0 1628--1637,
  2013.

\bibitem[Klucken et~al.(2013)Klucken, Barth, Kugler, Schlachetzki, Henze,
  Marxreiter, Kohl, Steidl, Hornegger, Eskofier, et~al.]{klucken2013unbiased}
Jochen Klucken, Jens Barth, Patrick Kugler, Johannes Schlachetzki, Thore Henze,
  Franz Marxreiter, Zacharias Kohl, Ralph Steidl, Joachim Hornegger, Bjoern
  Eskofier, et~al.
\newblock Unbiased and mobile gait analysis detects motor impairment in
  parkinson's disease.
\newblock \emph{PloS one}, 8\penalty0 (2), 2013.

\bibitem[Barrantes et~al.(2017)Barrantes, Egea, Rojas, Mart{\'\i}, Compta,
  Valldeoriola, Mezquita, Tolosa, and
  Valls-Sol{\`e}]{barrantes2017differential}
Sergi Barrantes, Antonio J~S{\'a}nchez Egea, Hern{\'a}n A~Gonz{\'a}lez Rojas,
  Maria~J Mart{\'\i}, Yaroslau Compta, Francesc Valldeoriola, Ester~Simo
  Mezquita, Eduard Tolosa, and Josep Valls-Sol{\`e}.
\newblock Differential diagnosis between parkinson's disease and essential
  tremor using the smartphone's accelerometer.
\newblock \emph{PloS one}, 12\penalty0 (8), 2017.

\bibitem[Lonini et~al.(2018)Lonini, Dai, Shawen, Simuni, Poon, Shimanovich,
  Daeschler, Ghaffari, Rogers, and Jayaraman]{lonini2018wearable}
Luca Lonini, Andrew Dai, Nicholas Shawen, Tanya Simuni, Cynthia Poon, Leo
  Shimanovich, Margaret Daeschler, Roozbeh Ghaffari, John~A Rogers, and Arun
  Jayaraman.
\newblock Wearable sensors for parkinson’s disease: which data are worth
  collecting for training symptom detection models.
\newblock \emph{NPJ digital medicine}, 1\penalty0 (1):\penalty0 1--8, 2018.

\bibitem[Butt et~al.(2017)Butt, Rovini, Esposito, Rossi, Maremmani, and
  Cavallo]{butt2017biomechanical}
Abdul~Haleem Butt, Erika Rovini, Dario Esposito, Giuseppe Rossi, Carlo
  Maremmani, and Filippo Cavallo.
\newblock Biomechanical parameter assessment for classification of
  parkinson’s disease on clinical scale.
\newblock \emph{International Journal of Distributed Sensor Networks},
  13\penalty0 (5):\penalty0 1550147717707417, 2017.

\bibitem[Zhan et~al.(2018)Zhan, Mohan, Tarolli, Schneider, Adams, Sharma,
  Elson, Spear, Glidden, Little, et~al.]{zhan2018using}
Andong Zhan, Srihari Mohan, Christopher Tarolli, Ruth~B Schneider, Jamie~L
  Adams, Saloni Sharma, Molly~J Elson, Kelsey~L Spear, Alistair~M Glidden,
  Max~A Little, et~al.
\newblock Using smartphones and machine learning to quantify parkinson disease
  severity: the mobile parkinson disease score.
\newblock \emph{JAMA neurology}, 75\penalty0 (7):\penalty0 876--880, 2018.

\bibitem[Battista and Romaniello(2018)]{battista2018novel}
Luigi Battista and Antonietta Romaniello.
\newblock A novel device for continuous monitoring of tremor and other motor
  symptoms.
\newblock \emph{Neurological Sciences}, 39\penalty0 (8):\penalty0 1333--1343,
  2018.

\bibitem[Heijmans et~al.(2019)Heijmans, Habets, Kuijf, Kubben, and
  Herff]{heijmans2019evaluation}
Margot Heijmans, Jeroen Habets, Mark Kuijf, Pieter Kubben, and Christian Herff.
\newblock Evaluation of parkinson’s disease at home: Predicting tremor from
  wearable sensors.
\newblock In \emph{2019 41st Annual International Conference of the IEEE
  Engineering in Medicine and Biology Society (EMBC)}, pages 584--587. IEEE,
  2019.

\bibitem[Hickey et~al.(2016)Hickey, Gunn, Alcock, Del~Din, Godfrey, Rochester,
  and Galna]{hickey2016validity}
Aodh{\'a}n Hickey, Eleanor Gunn, Lisa Alcock, Silvia Del~Din, Alan Godfrey,
  Lynn Rochester, and Brook Galna.
\newblock Validity of a wearable accelerometer to quantify gait in
  spinocerebellar ataxia type 6.
\newblock \emph{Physiological measurement}, 37\penalty0 (11):\penalty0 N105,
  2016.

\bibitem[LeMoyne et~al.(2016)LeMoyne, Heerinckx, Aranca, De~Jager, Zesiewicz,
  and Saal]{lemoyne2016wearable}
Robert LeMoyne, Frederic Heerinckx, Tanya Aranca, Robert De~Jager, Theresa
  Zesiewicz, and Harry~J Saal.
\newblock Wearable body and wireless inertial sensors for machine learning
  classification of gait for people with friedreich's ataxia.
\newblock In \emph{2016 IEEE 13th International Conference on Wearable and
  Implantable Body Sensor Networks (BSN)}, pages 147--151. IEEE, 2016.

\bibitem[Phan et~al.(2019{\natexlab{a}})Phan, Nguyen, Pathirana, Horne, Power,
  and Szmulewicz]{phan2019random}
Dung Phan, Nhan Nguyen, Pubudu~N Pathirana, Malcolm Horne, Laura Power, and
  David Szmulewicz.
\newblock A random forest approach for quantifying gait ataxia with truncal and
  peripheral measurements using multiple wearable sensors.
\newblock \emph{IEEE Sensors Journal}, 20\penalty0 (2):\penalty0 723--734,
  2019{\natexlab{a}}.

\bibitem[Phan et~al.(2019{\natexlab{b}})Phan, Nguyen, Pathirana, Horne, Power,
  and Szmulewicz]{phan2019quantitative}
Dung Phan, Nhan Nguyen, Pubudu~N Pathirana, Malcolm Horne, Laura Power, and
  David Szmulewicz.
\newblock Quantitative assessment of ataxic gait using inertial sensing at
  different walking speeds.
\newblock In \emph{2019 41st Annual International Conference of the IEEE
  Engineering in Medicine and Biology Society (EMBC)}, pages 4600--4603. IEEE,
  2019{\natexlab{b}}.

\bibitem[Ilg et~al.(2020)Ilg, Seemann, Giese, Trasch{\"u}tz, Sch{\"o}ls,
  Timmann, and Synofzik]{ilg2020real}
Winfried Ilg, Jens Seemann, Martin Giese, Andreas Trasch{\"u}tz, Ludger
  Sch{\"o}ls, Dagmar Timmann, and Matthis Synofzik.
\newblock Real-life gait assessment in degenerative cerebellar ataxia: Toward
  ecologically valid biomarkers.
\newblock \emph{Neurology}, 95\penalty0 (9):\penalty0 e1199--e1210, 2020.

\bibitem[Tran et~al.(2019)Tran, Pathirana, Horne, Power, and
  Szmulewicz]{tran2019automated}
Ha~Tran, Pubudu~N Pathirana, Malcolm Horne, Laura Power, and David~J
  Szmulewicz.
\newblock Automated evaluation of upper limb motor impairment of patient with
  cerebellar ataxia.
\newblock In \emph{2019 41st Annual International Conference of the IEEE
  Engineering in Medicine and Biology Society (EMBC)}, pages 6846--6849. IEEE,
  2019.

\bibitem[Nguyen et~al.(2018)Nguyen, Pathirana, Horne, Power, and
  Szmulewicz]{nguyen2018quantitative}
Khoa~D Nguyen, Pubudu~N Pathirana, Malcolm Horne, Laura Power, and David
  Szmulewicz.
\newblock Quantitative assessment of cerebellar ataxia with kinematic sensing
  during rhythmic tapping.
\newblock In \emph{2018 40th Annual International Conference of the IEEE
  Engineering in Medicine and Biology Society (EMBC)}, pages 1098--1101. IEEE,
  2018.

\bibitem[Gavriel et~al.(2015)Gavriel, Thomik, Louren{\c{c}}o, Nageshwaran,
  Athanasopoulos, Sylaidi, Festenstein, and Faisal]{gavriel2015towards}
Constantinos Gavriel, Andreas~AC Thomik, Pedro~Rente Louren{\c{c}}o, Sathiji
  Nageshwaran, Stavros Athanasopoulos, Anastasia Sylaidi, Richard Festenstein,
  and A~Aldo Faisal.
\newblock Towards neurobehavioral biomarkers for longitudinal monitoring of
  neurodegeneration with wearable body sensor networks.
\newblock In \emph{2015 7th International IEEE/EMBS Conference on Neural
  Engineering (NER)}, pages 348--351. IEEE, 2015.

\bibitem[Taniguchi et~al.(2011)Taniguchi, Hamahata, and
  Iwahashi]{taniguchi2011unsupervised}
Tadahiro Taniguchi, Keita Hamahata, and Naoto Iwahashi.
\newblock Unsupervised segmentation of human motion data using a sticky
  hierarchical dirichlet process-hidden markov model and minimal description
  length-based chunking method for imitation learning.
\newblock \emph{Advanced Robotics}, 25\penalty0 (17):\penalty0 2143--2172,
  2011.

\bibitem[Wiltschko et~al.(2015)Wiltschko, Johnson, Iurilli, Peterson, Katon,
  Pashkovski, Abraira, Adams, and Datta]{wiltschko2015mapping}
Alexander~B Wiltschko, Matthew~J Johnson, Giuliano Iurilli, Ralph~E Peterson,
  Jesse~M Katon, Stan~L Pashkovski, Victoria~E Abraira, Ryan~P Adams, and
  Sandeep~Robert Datta.
\newblock Mapping sub-second structure in mouse behavior.
\newblock \emph{Neuron}, 88\penalty0 (6):\penalty0 1121--1135, 2015.

\bibitem[Wiltschko et~al.(2020)Wiltschko, Tsukahara, Zeine, Anyoha, Gillis,
  Markowitz, Peterson, Katon, Johnson, and Datta]{wiltschko2020revealing}
Alexander~B Wiltschko, Tatsuya Tsukahara, Ayman Zeine, Rockwell Anyoha,
  Winthrop~F Gillis, Jeffrey~E Markowitz, Ralph~E Peterson, Jesse Katon,
  Matthew~J Johnson, and Sandeep~Robert Datta.
\newblock Revealing the structure of pharmacobehavioral space through motion
  sequencing.
\newblock \emph{Nature Neuroscience}, 23\penalty0 (11):\penalty0 1433--1443,
  2020.

\bibitem[Goetz et~al.(2008)Goetz, Tilley, Shaftman, Stebbins, Fahn,
  Martinez-Martin, Poewe, Sampaio, Stern, Dodel, et~al.]{goetz2008movement}
Christopher~G Goetz, Barbara~C Tilley, Stephanie~R Shaftman, Glenn~T Stebbins,
  Stanley Fahn, Pablo Martinez-Martin, Werner Poewe, Cristina Sampaio,
  Matthew~B Stern, Richard Dodel, et~al.
\newblock Movement disorder society-sponsored revision of the unified
  parkinson's disease rating scale (mds-updrs): scale presentation and
  clinimetric testing results.
\newblock \emph{Movement disorders: official journal of the Movement Disorder
  Society}, 23\penalty0 (15):\penalty0 2129--2170, 2008.

\bibitem[Lee et~al.(2019)Lee, Gommers, Waselewski, Wohlfahrt, and
  O’Leary]{lee2019pywavelets}
Gregory~R Lee, Ralf Gommers, Filip Waselewski, Kai Wohlfahrt, and Aaron
  O’Leary.
\newblock Pywavelets: A python package for wavelet analysis.
\newblock \emph{Journal of Open Source Software}, 4\penalty0 (36):\penalty0
  1237, 2019.

\bibitem[Daubechies et~al.(2011)Daubechies, Lu, and
  Wu]{daubechies2011synchrosqueezed}
Ingrid Daubechies, Jianfeng Lu, and Hau-Tieng Wu.
\newblock Synchrosqueezed wavelet transforms: An empirical mode
  decomposition-like tool.
\newblock \emph{Applied and computational harmonic analysis}, 30\penalty0
  (2):\penalty0 243--261, 2011.

\bibitem[MATLAB Wavelet Toolbox, The MathWorks, Natick, MA,
  USA()]{matlabwavelet}
MATLAB Wavelet Toolbox, The MathWorks, Natick, MA, USA.
\newblock Matlab wavelet toolbox, the mathworks, natick, ma, usa, R2018b.

\bibitem[Teh et~al.(2005)Teh, Jordan, Beal, and Blei]{teh2005sharing}
Yee~W Teh, Michael~I Jordan, Matthew~J Beal, and David~M Blei.
\newblock Sharing clusters among related groups: Hierarchical dirichlet
  processes.
\newblock In \emph{Advances in neural information processing systems}, pages
  1385--1392, 2005.

\bibitem[Chen et~al.(2004)Chen, Liaw, Breiman, et~al.]{chen2004using}
Chao Chen, Andy Liaw, Leo Breiman, et~al.
\newblock Using random forest to learn imbalanced data.
\newblock \emph{University of California, Berkeley}, 110\penalty0
  (1-12):\penalty0 24, 2004.

\bibitem[Lema{{\^i}}tre et~al.(2017)Lema{{\^i}}tre, Nogueira, and
  Aridas]{imblearn}
Guillaume Lema{{\^i}}tre, Fernando Nogueira, and Christos~K. Aridas.
\newblock Imbalanced-learn: A python toolbox to tackle the curse of imbalanced
  datasets in machine learning.
\newblock \emph{Journal of Machine Learning Research}, 18\penalty0
  (17):\penalty0 1--5, 2017.
\newblock URL \url{http://jmlr.org/papers/v18/16-365}.

\bibitem[Pedregosa et~al.(2011)Pedregosa, Varoquaux, Gramfort, Michel, Thirion,
  Grisel, Blondel, Prettenhofer, Weiss, Dubourg, Vanderplas, Passos,
  Cournapeau, Brucher, Perrot, and Duchesnay]{scikit-learn}
F.~Pedregosa, G.~Varoquaux, A.~Gramfort, V.~Michel, B.~Thirion, O.~Grisel,
  M.~Blondel, P.~Prettenhofer, R.~Weiss, V.~Dubourg, J.~Vanderplas, A.~Passos,
  D.~Cournapeau, M.~Brucher, M.~Perrot, and E.~Duchesnay.
\newblock Scikit-learn: Machine learning in {P}ython.
\newblock \emph{Journal of Machine Learning Research}, 12:\penalty0 2825--2830,
  2011.

\bibitem[Oung et~al.(2018)Oung, Muthusamy, Basah, Lee, and
  Vijean]{oung2018empirical}
Qi~Wei Oung, Hariharan Muthusamy, Shafriza~Nisha Basah, Hoileong Lee, and
  Vikneswaran Vijean.
\newblock Empirical wavelet transform based features for classification of
  parkinson’s disease severity.
\newblock \emph{Journal of medical systems}, 42\penalty0 (2):\penalty0 29,
  2018.

\end{thebibliography}

\end{document}